\documentclass[preprint,aps,superscriptaddress,pra]{revtex4-2}
\usepackage{graphicx}
\usepackage{dcolumn}
\usepackage{textcomp}
\usepackage{gensymb}

\begin{document}
\title{Ultrafast dynamics of ferroelectric polarization of NbOI\textsubscript{2} captured with femtosecond electron diffraction}
\author{Yibo Wang}
\email{yibowang@stanford.edu}
\affiliation{
Department of Physics and Astronomy, University of Nebraska-Lincoln, 855 N 16th Street, Lincoln, Nebraska 68588, USA
}
\affiliation{
Stanford PULSE Institute, SLAC National Accelerator Laboratory, 2575 Sand Hill Road, Menlo Park, California 94025, USA
}
\author{Md Sazzad Hossain}
\affiliation{
Department of Chemistry, University of Nebraska-Lincoln, 855 N 16th Street, Lincoln, Nebraska 68588, USA
}
\author{Tianlin Li}
\affiliation{
Department of Physics and Astronomy, University of Nebraska-Lincoln, 855 N 16th Street, Lincoln, Nebraska 68588, USA
}
\author{Yanwei Xiong}
\affiliation{
Department of Physics and Astronomy, University of Nebraska-Lincoln, 855 N 16th Street, Lincoln, Nebraska 68588, USA
}
\author{Cuong Le}
\affiliation{
Department of Physics and Astronomy, University of Nebraska-Lincoln, 855 N 16th Street, Lincoln, Nebraska 68588, USA
}
\author{Jesse Kuebler}
\affiliation{
Department of Mechanical and Materials Engineering, University of Nebraska-Lincoln, 855 N 16th Street, Lincoln, Nebraska 68588, USA
}
\author{Nina Raghavan}
\affiliation{
Department of Physics and Astronomy, University of Nebraska-Lincoln, 855 N 16th Street, Lincoln, Nebraska 68588, USA
}
\affiliation{
Department of Mechanical Engineering, University of Maryland, College Park, Maryland 20742, USA
}
\author{Lucia Fernandez-Ballester}
\affiliation{
Department of Mechanical and Materials Engineering, University of Nebraska-Lincoln, 855 N 16th Street, Lincoln, Nebraska 68588, USA
}
\author{Xia Hong}
\affiliation{
Department of Physics and Astronomy, University of Nebraska-Lincoln, 855 N 16th Street, Lincoln, Nebraska 68588, USA
}
\author{Alexander Sinitskii}
\affiliation{
Department of Chemistry, University of Nebraska-Lincoln, 855 N 16th Street, Lincoln, Nebraska 68588, USA
}
\author{Martin Centurion}
\email{martin.centurion@unl.edu}
\affiliation{
Department of Physics and Astronomy, University of Nebraska-Lincoln, 855 N 16th Street, Lincoln, Nebraska 68588, USA
}
\date{\today}
\begin{abstract}
Two-dimensional (2D) ferroelectric materials like NbOI\textsubscript{2} have garnered significant interest, yet their temporal response and synergetic interaction with light remain underexplored.
Previous studies on the polarization of oxide ferroelectrics have relied on time-resolved optical second harmonic generation or ultrafast X-ray scattering. 
Here, we probe the laser-induced polarization dynamics of 2D NbOI\textsubscript{2} nanocrystals using ultrafast transmission electron diffraction and deflectometry.
The deflection of the electron pulses is directly sensitive to the changes in the polarization, while the diffraction signal captures the structural evolution.
Excited with a UV laser pulse, the polarization of NbOI\textsubscript{2} is initially suppressed for two picoseconds, then it recovers and overshoots, leading to a transiently enhanced polarization persisting for over 200 ps.
This recovery coincides with coherent acoustic phonon generation, triggering a piezoresponse in the NbOI\textsubscript{2} nanocrystals.
Our results offer a new method for sensing the ferroelectric order parameter in femtosecond time scales.
\end{abstract}
\maketitle
\newpage

Optically tunable ferroelectrics are highly desirable in photonic integrated circuits.
Understanding their dynamic response to light and energy dissipation pathways is critical for their application in optoelectronics, ultrasonic devices, and next-generation wireless communications, which leads to intense research on the ultrafast interactions between ferroelectric materials and optical stimuli 
\cite{daranciang2012ultrafast,mankowsky2017ultrafast,li2018optical,zhang2021probing,chen2022deterministic,luo2023ultrafast}.
Such optical tuning (a suppression or reversal of the ferroelectric polarization) can be achieved through THz pulses \cite{mankowsky2017ultrafast,grishunin2017thz,morimoto2017terahertz}, or a laser pulse of shorter wavelength \cite{chen2015ferroelectric,lian2019indirect,krapivin2022ultrafast,xu2022femtosecond}. 
In addition, domain dynamics could be introduced with femtosecond laser pulses \cite{chen2015ferroelectric,morimoto2017terahertz,park2022light,guzelturk2023sub}.
However, past studies have focused on oxide ferroelectrics, while the interaction between light and 2D van der Waals (vdW) ferroelectrics, which are key to the high flexibility and energy-efficient microelectronics \cite{ryu2020empowering}, remains scarcely explored.
The two-dimensional ferroelectric NbOI\textsubscript{2} has drawn significant interest due to its ferroelectricity at room temperature and significant terahertz emission \cite{jia2019,fang2021,chen2021,wu2022,cui2023,pan2023,handa20242d,huang2025coupling,subedi2025colossal}. 
It also exhibits the largest piezoelectric coefficients in the class of vdW ferroelectrics \cite{wu2022}.
Static investigations of this material using second harmonic generation (SHG) show that the polarization of NbOI\textsubscript{2} crystal is highly tunable by mechanical stress \cite{abdelwahab2022,ye2023manipulation,fu2024emission,fu2023manipulating}.
Unlike mechanical strain induced by sample wrinkles or high pressure, the optical strain induced by ultrafast optical stimulation is a transient phenomenon, and its dynamical effect on the sample can only be accessed by time-resolved techniques.
In this work, 2D NbOI\textsubscript{2} nanocrystals are excited by UV laser pulses, and the ferroelectric polarization dynamics are captured with femtosecond resolution using ultrafast transmission electron diffraction (UTED).
\par
Typically, the ultrafast nonequilibrium measurements of the ferroelectric order parameter and domain were performed with time-resolved SHG spectroscopy \cite{mankowsky2017ultrafast} and X-ray scattering \cite{krapivin2022ultrafast}.
Electrons, on the other hand, are also effective probes for ferroelectric polarization, as the Coulomb force between polarized crystals and electrons induces a detectable deflection in the electron beam \cite{lichte2002ferroelectric,shibata2012differential,polking2012ferroelectric,maclaren2015origin,taplin2016low,teodorescu2017low}.
In addition, high-energy electrons reveal the structural information of the samples in real and momentum space with minimal energy exchange, causing significantly less damage compared to other probes.
Here, we demonstrate the first ultrafast in-situ measurement of the polarization of ferroelectrics using pulsed electrons as a probe.
By tracking the deflection angle of the electron pulse, we achieved a temporally resolved characterization of the sample's polarization.
We show that an ultrafast UV excitation of NbOI\textsubscript{2} leads to both a rapid suppression of polarization in the first 2 ps and an enhancement to the ferroelectric order parameter after 20 ps.
In addition to the polarization dynamics, the electron diffraction patterns reveal the efficient generation of coherent acoustic phonons.
Using short electron pulses as the probe, the lattice strain and polarization are captured simultaneously and independently, giving an unprecedented view of the interplay of the thermal, stress/strain, and polarization aspects of the ferroelectric sample.

\section{Results}

\begin{figure}
    \centering
    \includegraphics[width=6 in]{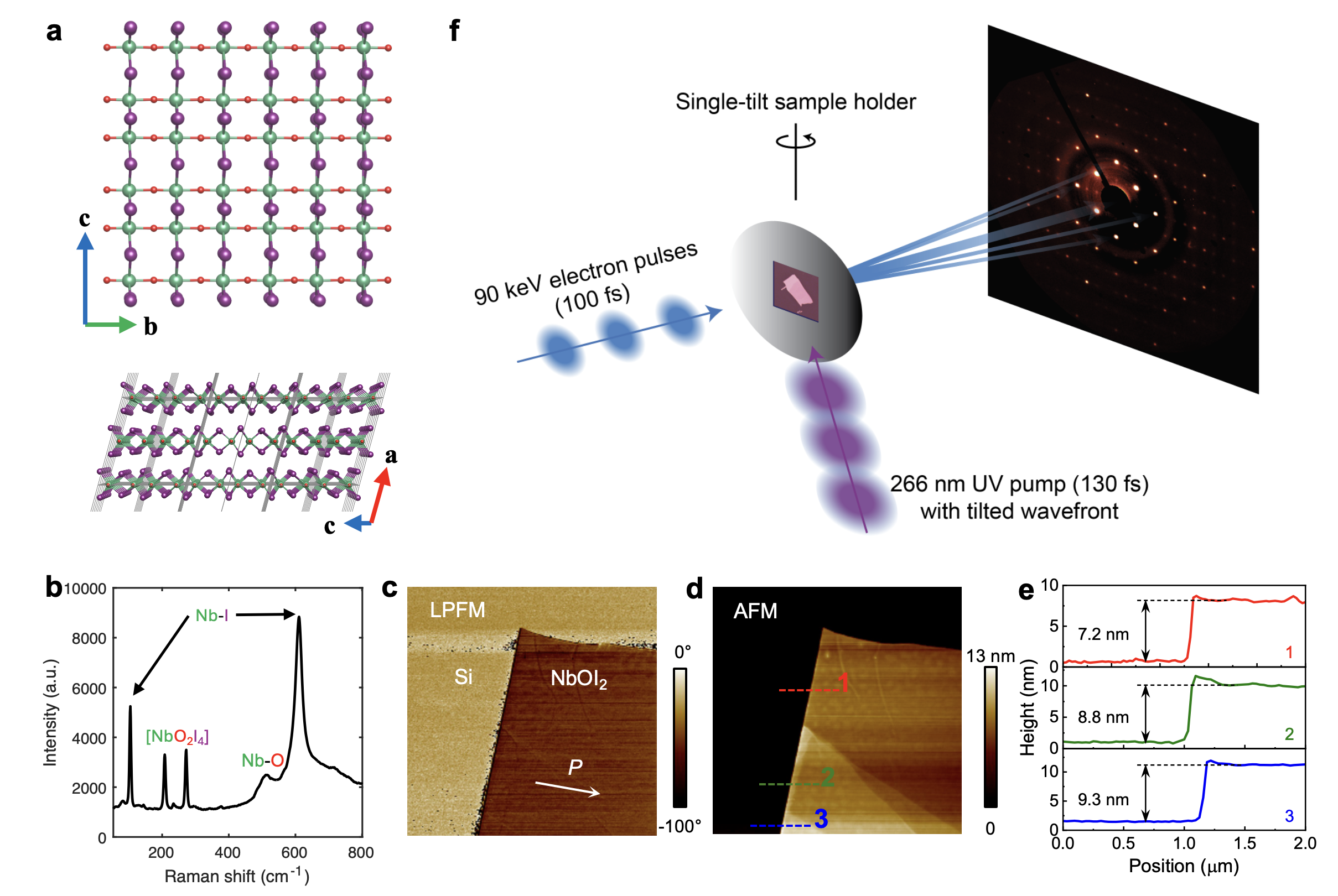}
    \caption{
    \textbf{a}, A top view of one layer of the NbOI\textsubscript{2} nanosheet (top) and a side view showing its monoclinic crystal lattice (bottom).
    \textbf{b}, Raman spectrum of bulk NbOI\textsubscript{2}.
    The 105 cm$^{-1}$ and 610 cm$^{-1}$ peaks correspond to symmetric bending and asymmetric stretching of the Nb-I bond, respectively.
    Peaks at 208 cm$^{-1}$ and 273 cm$^{-1}$ arise from [NbO\textsubscript{2}I\textsubscript{4}] octahedron vibrations, while the 514 cm$^{-1}$ peak reflects symmetric Nb-O bending.
    \textbf{c}, LPFM phase image of the NbOI\textsubscript{2} flake on Si with scanning angle of 10°, revealing a uniform in-plane polarization.
    \textbf{d}, AFM topography image taken on an NbOI\textsubscript{2} flake on Si wafer, with \textbf{e} the height profiles along the dashed lines.
    \textbf{f}, An illustration of the experimental setup. 
    The laser pump with 266 nm wavelength is incident on the sample at an angle of 58\degree.
    A 90 keV electron probe pulse of approximately 150 fs duration is scattered by the sample. 
    }
    \label{fig1}
\end{figure}

\subsection{Characterization of NbOI\textsubscript{2} nanocrystals}

Figure~\ref{fig1}a shows the crystal lattice of NbOI\textsubscript{2}.
The lattice constants of NbOI\textsubscript{2} are a = 15.18 \AA, b = 3.92 \AA, c = 7.52 \AA, and the angle $\beta$ between lattice vectors \textbf{\textit{a}} and \textbf{\textit{c}} is 105.50\degree\ \cite{fang2021}. 
Bulk NbOI\textsubscript{2} samples were grown using the chemical vapor transport (CVT) method (Methods).
Figure~\ref{fig1}b shows the Raman spectroscopy of bulk NbOI\textsubscript{2}, which revealed five distinct peaks at 105, 208, 273, 514, and 610 cm$^{-1}$, in agreement with the previously reported values \cite{fang2021}.
The thin samples used in UTED were mechanically exfoliated and transferred onto an amorphous Si\textsubscript{3}N\textsubscript{4} thin film.
As shown in Fig.~\ref{fig1}c, the piezoresponse force microscopy (PFM) phase image confirms a uniform in-plane polarization in the exfoliated sample flake.
Atomic force microscopy (AFM) was used to determine the sample morphology.
Figure~\ref{fig1}d shows the AFM topography image of an NbOI\textsubscript{2} flake whose thickness ranges from 7.2 nm to 9.3 nm (Fig.~\ref{fig1}e).
The data presented in this work was primarily obtained from two samples with thicknesses of 39 nm and 7 nm, both with a lateral size of around 100 $\mu$m.
The heat capacity of NbOI\textsubscript{2} crystals was measured using a differential scanning calorimeter (DSC) to be $0.27\pm0.04$ J/(gK) at 20 \degree C and $0.31\pm0.05$ J/(gK) at 80 \degree C.
The details of the PFM and DSC measurements can be found in supplementary materials (Sec.~\ref{Sec:S2}).

\begin{figure}
    \centering
    \includegraphics[width=4.5 in]{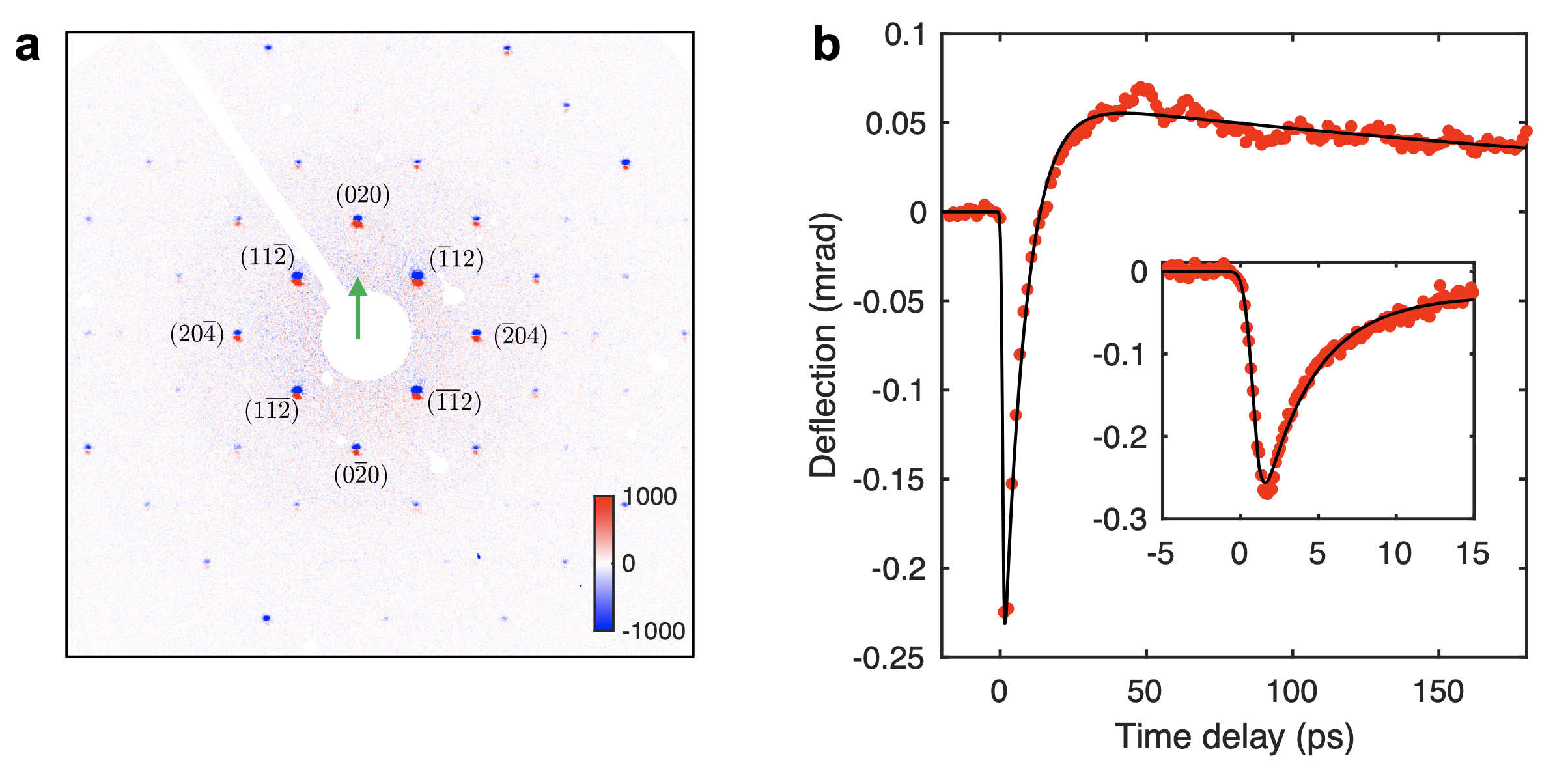}
    \caption{
    \textbf{a}, A differential image between the electron diffraction patterns taken at 2 ps after laser excitation and before the arrival of laser pulses. 
    The green arrows mark the polar \textbf{\textit{b\textsuperscript{*}}}-axis. 
    \textbf{b}, Averaged time-resolved deflection of the low-order \{11$\overline{2}$\}, \{020\} and \{20$\overline{4}$\} Bragg peaks. 
    }
    \label{fig1_S}
\end{figure}

\subsection{Transient deflection of pulsed electron beam}

Figure~\ref{fig1}f illustrates the schematic of the UTED geometry.
The samples, driven out of equilibrium using a 266 nm UV laser pulse with a fluence ranging from 0.5 to 1.4 mJ/cm\textsuperscript{2}, are probed with 90 keV electron pulses of approximately 150 fs duration (full width at half maximum, FWHM) at a precisely controlled time delay.
Upon the laser excitation of the 39 nm sample, we observed a uniform shift of all diffraction orders in the same direction.
Figure~\ref{fig1_S}a shows the difference between the diffraction patterns taken at a 2 ps time delay and before the laser excitation, where the pixels with increased counts are displayed in red and those with decreased counts in blue.
The differential image represents a uniform downward deflection of the entire diffraction pattern, and all Bragg peaks are deflected by the same angle on the detector.
Such deflection occurs along the polar \textbf{\textit{b}}-axis, which is marked by a green arrow in Fig.~\ref{fig1_S}a. 
The time-resolved movement averaged over the eight brightest diffraction orders is plotted in Fig.~\ref{fig1_S}b.
A zoomed-in scan with finer time steps over a shorter range is shown in the inset.
In less than 2 ps, the deflection of all Bragg orders reaches the maximum, followed by a recovery in the next 20 ps. 
The deflection changes direction and continues to increase until around 50 ps.
Subsequently, a slower relaxation extends over hundreds of picoseconds.
\par
We attribute the time-dependent deflection of the Bragg peaks depicted in Fig.~\ref{fig1_S}b to a change in the polarization of the sample.
The movement is independent of the diffraction order, ruling out the possibility of it being driven by changes in the lattice parameters.
Additionally, the movement is along the polar axis of the sample, along which a spontaneous electric field exists. 
The shifting of Bragg peaks results from the time-dependent deflection of the electron beam within the sample, induced by an evolving electric field inside the polarised crystal. 

\begin{figure}
    \centering
    \includegraphics[width=3.75 in]{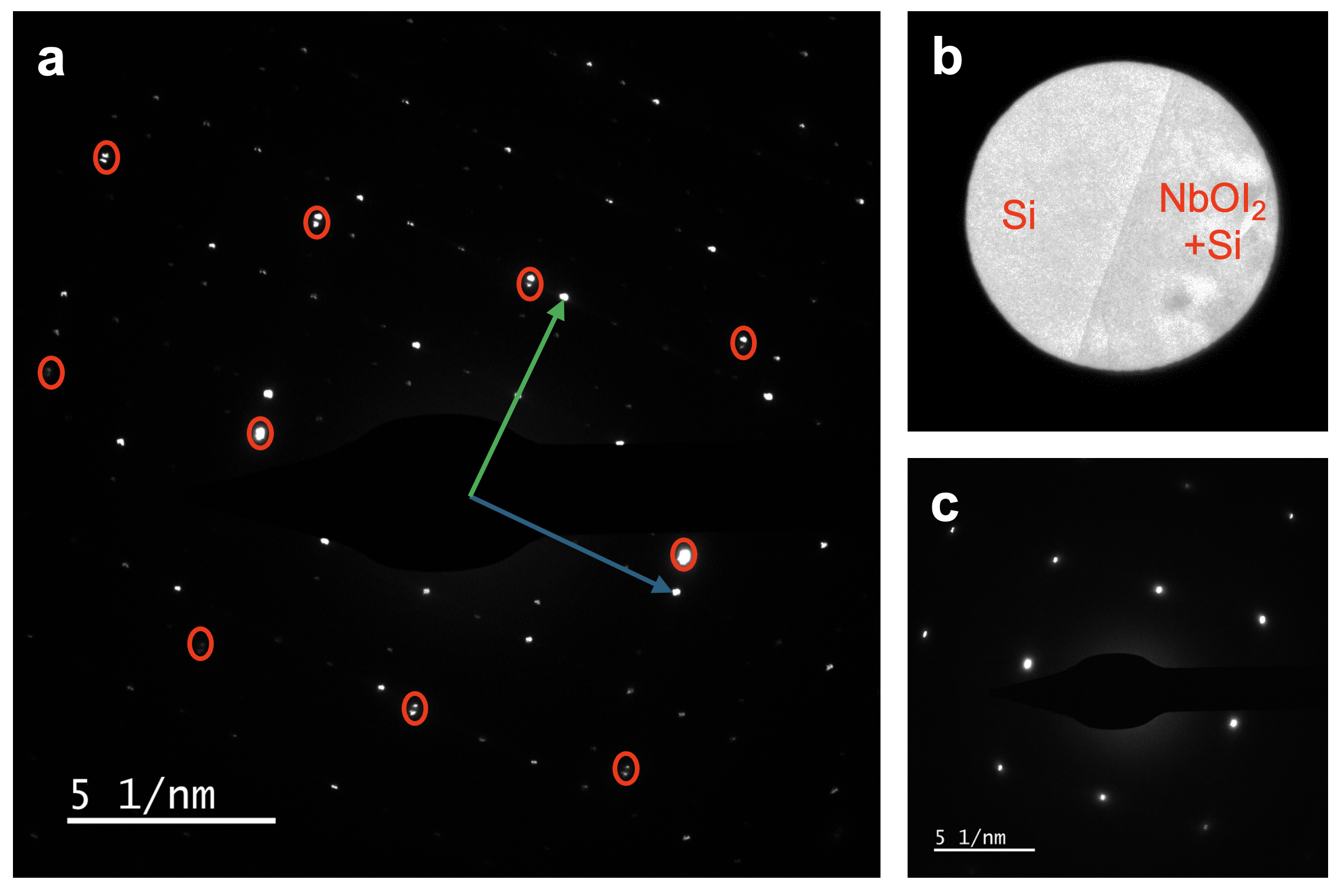}
    \caption{
    \textbf{a}, An SAED pattern consisting of Bragg peaks from both the NbOI\textsubscript{2} sample and the single-crystal silicon substrate. 
    The green and blue arrows mark the polar and non-polar axes, respectively. 
    The diffraction from silicon is circled in red and split along the polar axis. 
    \textbf{b}, The selected area where diffraction pattern in \textbf{a} is taken. 
    \textbf{c}, A reference SAED pattern taken from the silicon substrate only. 
    }
    \label{fig2}
\end{figure}

\par
To verify that the electron beam is deflected by the ferroelectric polarization, we performed selected area electron diffraction (SAED) using a transmission electron microscopy (Fig.~\ref{fig2}a).
The sample consists of a 10-nm-thick single-crystal NbOI\textsubscript{2} flake, which is deposited on top of a single-crystal silicon substrate of 15 nm thickness.
The transverse size of the sample is small compared to the substrate.
Thus, we can compare the diffraction from the sample and substrate together with the diffraction from the substrate only.
Electrons diffracted through the polarized crystal are expected to acquire an additional deflection compared to those traversing only the substrate.
We designed a selected area consisting of both the deflected and undeflected electrons, which is shown in Fig.~\ref{fig2}b, where half of the electron beam interacts with both the NbOI\textsubscript{2} sample and the silicon substrate, while the other half interacts only with the silicon substrate.
The resulting SAED pattern displayed in Fig.~\ref{fig2}a suggests that the Bragg peaks of the silicon thin film (circled in red) are split along the polar axis by 0.24 mrad in this diffraction geometry.
In contrast, as shown in Fig.~\ref{fig2}c, selecting a region containing only the silicon substrate results in unsplit Bragg peaks.
The disparity in the splitting of silicon Bragg peaks, with and without the NbOI\textsubscript{2} sample, clearly demonstrates the polarized crystal deflects a keV electron beam by a measurable angle. 
A quantitative analysis (Sec.~\ref{Sec:S1}) of the deflection angle gives an intrinsic electric field of $8\times10^9$ V/m, matching the intrinsic field of $9\times10^9$ V/m, calculated from the polarization and dielectric constants of NbOI\textsubscript{2} \cite{wu2022}. 
\par
Domain dynamics are confirmed to have no influence on the experimental results based on three reasons.
First, the PFM measurements performed on several freshly exfoliated NbOI\textsubscript{2} samples with lateral size on the order of 100 $\mu$m show a uniform polarization with no domain walls (Sec.~\ref{Sec:S2}).
Second, in the UTED experiments, the pump fluence is capped so that the thermalized temperature of the samples is below the Curie temperature of 189 \degree C \cite{wu2022}.
Third, the effect of domain structures in NbOI\textsubscript{2} samples would be to split or elongate the Bragg peaks along the polar axis due to the opposite directions of electric fields.
Such an effect was not observed. 

\subsection{Ultrafast suppression of ferroelectric order}

\begin{figure}
    \centering
    \includegraphics[width=5.5 in]{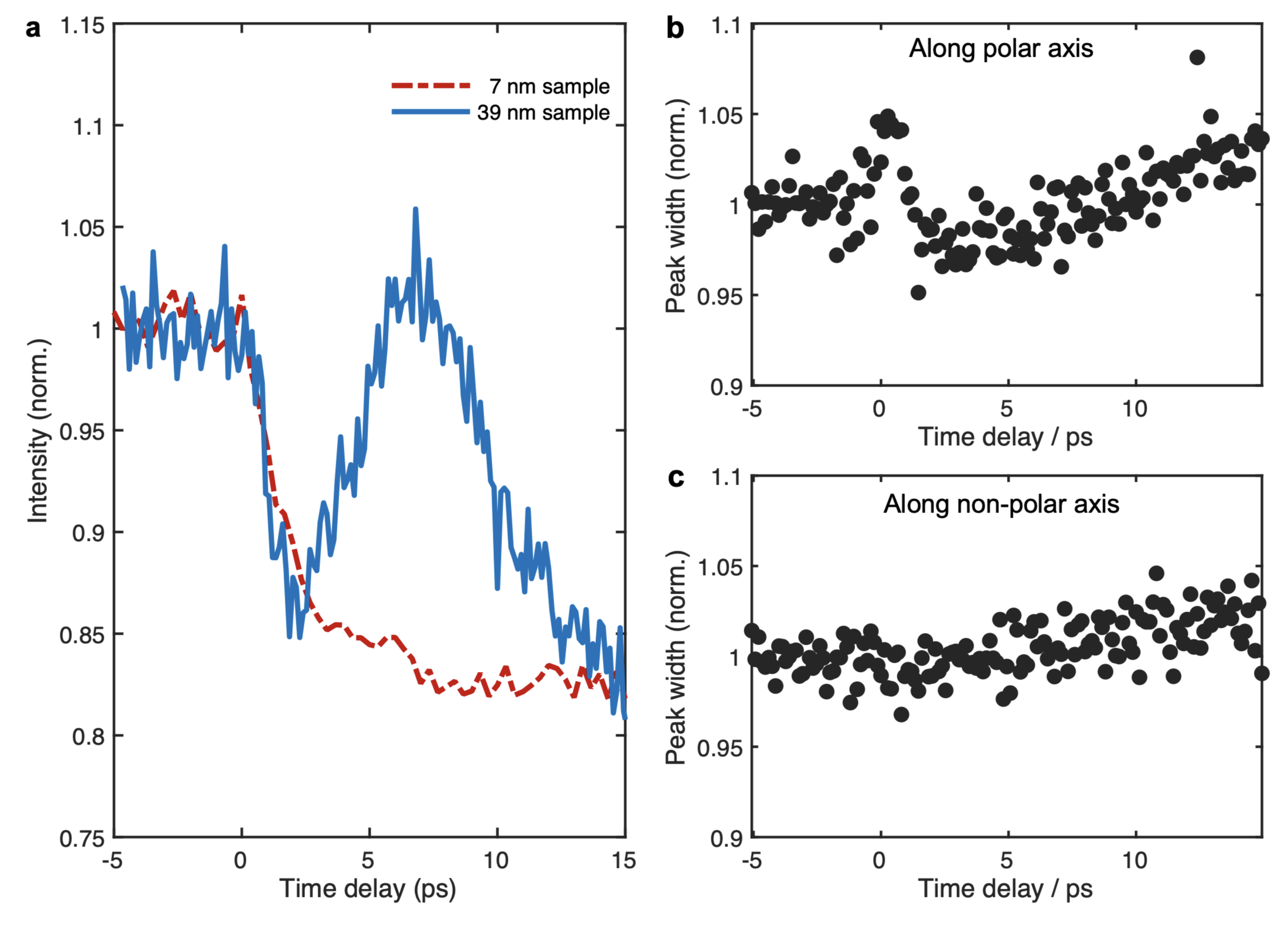}
    \caption{
    \textbf{a}, The normalized Bragg peak intensity of the (020) diffraction order from a 7-nm sample (red dashed line) and a 39-nm sample (blue line). 
    \textbf{b}, The normalized peak width along the polarized axis taken from the 39-nm sample. 
    \textbf{c}, The normalized peak width along the non-polar axis taken from the 39-nm sample. 
    A broadening of the Bragg peak occurs in the first picosecond along the polar axis, which is absent along the non-polar axis.
    }
    \label{fig3}
\end{figure}

The changes in polarization, as seen in Fig.~\ref{fig1_S}b, have three distinct time scales:
1. From excitation up to 2 ps, the diffraction orders move and reach a maximum deflection angle.
2. Over the next 50 ps, the deflection angle recovers towards zero and overshoots in the opposite direction.
3. Subsequently, a slower relaxation on the timescale of more than a hundred picoseconds occurs.
The deflection angle of the electron pulse is phenomenologically described as:
\begin{equation}
    \Delta S = A_1\,\textrm{erfc}\,(\frac{t-t_1}{\tau_1}) + A_2\,\textrm{exp}\,(-\frac{t-t_2}{\tau_2})\cdot \textrm{exp}\,(-\frac{t-t_3}{\tau_3})+C,
    \label{equ1}
\end{equation}
where a complementary error function with time constant $\tau_1$ represents the rapid suppression of polarization; two exponential terms with time constants $\tau_2$ and $\tau_3$ describe the recovery and the slow relaxation.
A fit of equation~\ref{equ1} to the data (Fig.~\ref{fig1_S}b) returns the time constants $\tau_1 = 0.679$ ps, $\tau_2 = 7.75$ ps and $\tau_3 = 299$ ps.
\par
In the first two picoseconds, the polarization of the samples is rapidly suppressed by two factors: a rapid lattice temperature rise and carrier charge screening.
As shown in Fig.~\ref{fig3}a, the ultrafast lattice heating is evidenced by the intensity reduction of the (020) diffraction order, observed in the 39 nm and 7 nm thick samples, corresponding to inhomogeneous and homogeneous heating, respectively.
Based on a reported imaginary dielectric coefficient $\varepsilon_2$ of 6.4 for 266 nm UV laser \cite{abdelwahab2022}, we estimate the absorption length of the pump pulse to be 10 nm. 
In the 7-nm thin sample, the dynamics resemble the Debye-Waller effect, and the rise in temperature is mostly uniform across its thickness.
The diffraction intensity continuously decreases until the lattice temperature reaches the maximum at around 7 ps.
In the thick sample, the Bragg intensity also sees a rapid suppression in the first 2 ps, after which the dynamics are dominated by the strain wave.
In both cases, the temperature of the lattice rapidly increases in the first 2 ps, which matches the timescale for the initial change in the polarization depicted in Fig.~\ref{fig1_S}b.
\par
Another contributing factor to the reduction of polarization is the charge screening effect. 
The absorption of the femtosecond pump pulse generates hot carriers, leading to the screening of the internal polarization field \cite{daranciang2012ultrafast,luo2023ultrafast} and the emission of coherent acoustic phonons.
Figure ~\ref{fig3}b, c show the normalized FWHM of the (020) diffraction order in the 39 nm sample along the polar and non-polar axes, respectively.
A broadening of the Bragg peak along the polar axis is found in the first picosecond after laser excitation, which is absent along the non-polar axis. 
The anisotropic broadening of the Bragg peak suggests it is unrelated to lattice heating.
The broadening of the Bragg peak along the polar axis results from an uneven screening of the internal polarization field along the thickness dimension of the sample.
In the 39 nm sample, a gradient of electron temperature is created after the absorption of the laser pump pulse, leading to a depth-dependent dampening of the intrinsic electric field in the sample.
As a result, a minor smearing of the electron beam is caused along the polar axis.
At maximum, the smearing reaches 0.01 mrad in real space, which is much smaller than the deflection (0.27 mrad at maximum) and occurs only within the first picosecond, indicating it cannot result from domain dynamics.
\par
In the following picosecond, the smearing of the internal polarization field vanishes, which is likely a result of the thermalization of the carrier system via electron-electron scattering across layers of the 2D sample.
Coherent strain waves form and dominate the dynamics after 2 ps, exerting a significant influence on the sample's polarization due to its strong piezoelectric response.

\subsection{Laser-induced strain dynamics}

\begin{figure}
    \centering
    \includegraphics[width=5.5 in]{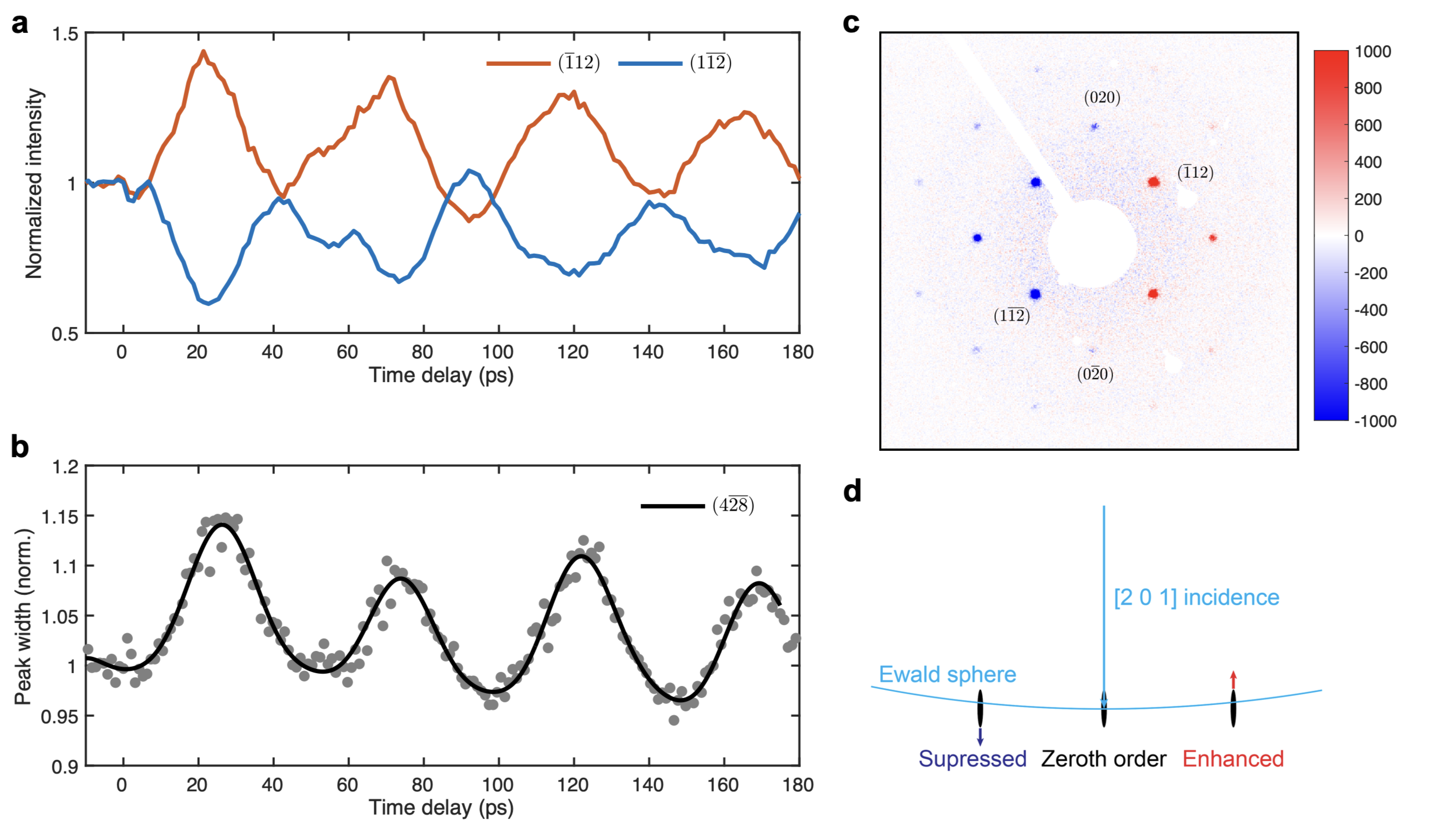}
    \caption{
    \textbf{a}, The diffraction intensity of the ($\overline{1}12$) and ($1\overline{12}$) orders. 
    \textbf{b}, The temporal evolution of the FWHM of $(4\overline{28})$ diffraction order along the polar axis. 
    \textbf{c}, A differential image showing the diffraction pattern taken at 25 ps delay time, subtracting the diffraction pattern taken before the arrival of the laser pulse.
    \textbf{d}, The Ewald sphere of the 90 keV electrons.
    A seesaw motion in the reciprocal lattice causes the antisymmetric modulation to the Bragg condition of a Friedel pair. 
    }
    \label{fig4}
\end{figure}

Coherent acoustic waves are launched along the thickness of the sample due to the transiently increased thermal stress in the electronic and lattice systems, charge screening, and inverse piezoelectric response \cite{park2005,harb2009,lejman2014giant,ruello2015physical,wei2017dynamic,qian2020,zong2023}.
As depicted in Fig.~\ref{fig1_S}b and Fig.~\ref{fig3}, the effects of acoustic waves on the diffraction pattern are clearly observable after 2 ps.
The polarization changes from a rapid reduction to a fast recovery;
the thermal suppression of Bragg diffraction intensity is overtaken by the modulation from lattice deformation in the presence of the acoustic wave;
the Bragg peak width gradually increases after 2 ps, as part of a longer oscillatory behavior induced by the acoustic wave, which has a substantially longer period.
\par
The interlayer lattice motions induced by the acoustic waves are examined by the time-resolved intensity and FWHM of the Bragg peaks from the 39-nm-thick sample, shown in Fig.~\ref{fig4}a and b.
The diffraction intensities of a Friedel pair consisting of ($\overline{1}12$) and ($1\overline{12}$) diffraction orders oscillate at 20 GHz with opposite phases (Fig.~\ref{fig4}a).
The anti-symmetric phase distribution about the polar axis is displayed in Fig.~\ref{fig4}c.
This phenomenon is attributed to the time-varying, antisymmetric excitation error in the diffraction geometry, caused by the rotation of the reciprocal lattice.
Figure.~\ref{fig4}d illustrates such wobbling motions of the reciprocal lattice.
In momentum space, a seesaw motion occurs in the reciprocal lattice due to the shear acoustic wave, causing the diffraction orders on the left to deviate from the Bragg condition while those on the right shift towards it.
The seesaw motion of the reciprocal lattice can be further demonstrated by the evolution of the Bragg peak width, which oscillates on the same 20 GHz frequency.
Figure~\ref{fig4}b shows the FWHM of the $(4\overline{28})$ diffraction order, where the larger wavevector causes a significant deviation from its Bragg condition, resulting in weaker diffraction intensity under static conditions.
Under the influence of the shear wave, the diffraction geometry for the $(4\overline{28})$ diffraction order periodically gets closer to and away from the Bragg condition, with its FWHM changing accordingly. 
\par
In the thicker sample, where the laser's absorption length is significantly smaller than the crystal's thickness, the acoustic waves are launched as traveling blast waves along the sample's thickness dimension.
The wavefront of the acoustic wave is reflected at the bottom of the sample, and the reflected wave propagates upward.
The oscillation frequency is determined by the sound velocity and the thickness of the sample: $f=v/2D$, where $v$ is the interlayer sound velocity and $D$ is the sample thickness.
Further discussions of the modes of acoustic waves may be found in the supplementary materials (Sec.~\ref{Sec:S4}).
The traveling acoustic wave cools the hotter lattice layers by transferring heat across the longitudinal dimension of the sample, which leads to a fast recovery of the average polarization.
As depicted in Fig.~\ref{fig1_S}b, the thermal relaxation of the thermalized NbOI\textsubscript{2} lattice happens subsequently.
It spans out into hundreds of picoseconds, which fall into the timescale of thermal relaxation via diffusion into the substrate.
During this time, the oscillations in the intensity and width of Bragg peaks continue with a decreasing amplitude.

\section{Discussion}

The 266 nm UV laser pulses effectively excite the NbOI\textsubscript{2} samples above its indirect bandgap of $E_g = 2.24\,\mathrm{eV}$ \cite{abdelwahab2022}.
At different delay times, the polarization of the NbOI\textsubscript{2} samples is affected by different mechanisms.
Initially after laser excitation, the polarization is suppressed due to the dynamic effects of nonthermal heating of carriers and lattice at the vacuum-sample interface.
In a thermalizing lattice, the interfacial dynamics compete with the dynamics in the bulk crystal, where a coherent acoustic wave conducts the heat and applies inhomogeneous strain to the lattice.
The suppression reaches the maximum at 2 ps, indicating the effective dissipation of heat, marking the dynamics in the bulk crystal as the dominating factor.
After 20 ps, the polarization is enhanced in a thermalized lattice due to the persistent acoustic waves and its piezoelectric effects.
\par
The inhomogeneous lattice strain plays a key role in the enhancement of the polarization.
However, we lack direct evidence that a normal stress/strain exists along the polar axis of NbOI\textsubscript{2} samples, as the wavevector of the strain waves is along the thickness dimension.
The transverse and longitudinal modes of the acoustic wave give rise to $\sigma_5$ (shear stress in \textbf{\textit{ac}} plane) and $\sigma_1$ (normal stress along \textbf{\textit{a}}) elements in the stress tensor, respectively, which are insufficient to account for the changed polarization along the polar axis, given the major element in the piezoelectric tensor is $\varepsilon_{22}$.
We hypothesize that the enhanced polarization is driven by some of the following mechanisms.
First, the lattice strain in the longitudinal direction will trigger a local response in the lateral direction with an opposite sign due to the Poisson effect \cite{mante2018directional}.
Therefore, the compressive strain applied in the leading edge of the traveling strain wave would lead to a tensile strain along the polar axis.
Second, it is reported in transition metal dichalcogenides that out-of-plane acoustic waves are coupled to in-plane ones \cite{cremons2017defect}, which would also lead to lattice strain along the polar axis if the same phenomena also exist in NbOI\textsubscript{2}.
The detection of such coupling will require the bright-field imaging of ultrafast electron microscopy.
Third, flexoelectricity that emerges along with lattice strain gradient may also contribute to the fast enhancement of the polarization \cite{zubko2013flexoelectric}.
\par
In conclusion, we captured the ultrafast polarization changes and lattice dynamics of NbOI\textsubscript{2} crystals with UTED.
By monitoring the deflection of the electron diffraction pattern, we demonstrate the first time-resolved measurements of the polarization of ferroelectric samples using pulsed electrons.
We found that the polarization of NbOI\textsubscript{2} is enhanced for more than 200 ps after laser excitation, resulting from the highly piezoelectric response of NbOI\textsubscript{2}.
Moreover, coherent acoustic phonons are generated with great efficiency, which is evidenced by large GHz oscillations in the intensity and width of the Bragg peaks.
Additionally, evidence of an inhomogeneous carrier screening is found in the transient broadening of Bragg peaks along the polar axis.
Using ultrashort pulsed electrons, dynamical information of the carrier system, lattice system, and electrical properties of NbOI\textsubscript{2} are simultaneously obtained.
This temporally resolved characterization of the ultrafast response of NbOI\textsubscript{2} to the laser excitation provides insights into the optical control of its strain and polarization, as well as providing a new paradigm for sensing ferroelectric polarization in real space and time.
\clearpage
\begin{center}
\textbf{\large Methods}
\end{center}

\setcounter{section}{0}

\section{Experimental setup}

The UTED instrument was described in detail in a previous publication \cite{wang2023ultrafast}. 
The optical branches used for pumping the sample and generating the electron pulses are derived from the output of a Ti:Sapphire laser system (Coherent Astrella), which operates at a variable repetition rate of up to 1 kHz at a wavelength of 800 nm, subsequently harmonically tripled to 266 nm.
In this study, a 200 Hz repetition rate is used.
The UV pump pulses of approximately 130 fs duration and 187 $\mu$m diameter are delivered on the sample, providing photoexcitation. 
The probe branch shines UV pulses on a copper cathode to generate coherent photoelectron pulses, which are accelerated by a 90 kV electrostatic field before being focused by magnetic lenses.
The electron pulses interact with the sample and diffract in a transmission geometry. 
One translation stage is employed in the pump branch to control the time of arrival difference between the pump and probe pulses. 
\par
We tilt the energy front of the pump pulses to achieve simultaneous excitation of the sample. 
In order to do so, we placed an optical diffraction grating in the pump branch. 
By designing the incident angle of the laser pulse onto the grating and selecting the correct diffraction order, we achieved a 58\degree\ titled energy front of the diffracted laser pulse. 
Together with the 58\degree\ incidence of the pump pulses, the optical pump simultaneously excites the sample across the lateral direction and travels at the same velocity as the electron pulse along the longitudinal direction \cite{zhang2014tilted}. 

\section{Sample preparation}

The bulk crystals of NbOI\textsubscript{2} are prepared using the CVD method.
The starting materials Nb (powder), Nb\textsubscript{2}O\textsubscript{5} (powder), and I\textsubscript{2} (crystal) were mixed to form a stoichiometric ratio of Nb:O:I = 1:1:2 and sealed in a quartz ampule under vacuum (10\textsuperscript{-5} mbar). 
Excess iodine was added as a transport agent. 
The sealed ampule was placed in a horizontal dual-zone furnace and brought to 700\degree C slowly (at a rate of 1\degree C/min). 
The ampule was held for 7 days and cooled at room temperature naturally. 
The crystals were collected by opening the ampule in the inert conditions of the N\textsubscript{2}-filled glove box and soaking in IPA to remove excess iodine. 
The harvested samples were shiny crystals. 
\par
We mechanically exfoliated the bulk NbOI\textsubscript{2} into thin flakes with scotch tape and stamped them onto a silicone gel film (Gel-Pak Gel-Film PF-40-X4). 
We identified sample flakes with a thickness suitable for UTED experiments with optical microscopy, and then the sample flakes were transferred onto a SiN TEM window via high-temperature dry transfer technique \cite{liang2021two,bie2021versatile}. 
A home-built transfer stage was adapted for horizontal alignment and fine vertical adjustments.
The selected NbOI\textsubscript{2} flake was kept aligned to the desired location as the gel film approached the TEM grid.
After the gel film contacted the TEM window, gentle pressure was applied, and the transfer stage was heated to 80\degree C for 300 seconds for better adhesion. 
Then, the gel film was gently released from the TEM window, with the desired sample flake attached to the SiN window. 



\section{Atomic/piezoresponse force microscope}

The surface morphology and ferroelectric domain imaging were carried out using a Bruker Multimode 8 AFM.
The AFM and lateral piezoresponse force microscopy (LPFM) studies were performed using Pt/Ir-coated probes (Bruker SCM-PIT-V2). 
After the UTED experiments were done, the SiN window that carries the sample flake was attached to a free-standing polypropylene carbonate (PPC) film and then detached from the TEM grid. 
Subsequently, the SiN thin film was transferred onto a piece of silicon wafer before the PPC was removed with anisole.  
The thickness of the NbOI\textsubscript{2} samples was then measured with the AFM. 
During the process, the areas with wrinkles were avoided.
\par
The PFM measurements were taken at a drive frequency of 700 kHz with a bias voltage of 400 mV. 
The scanning angle $\phi$ is defined as the relative angle between the scanning direction and the in-plane polarization. 
For angle-dependent measurements, we fixed the cantilever scan direction and rotated the sample. 
The in-plane polar axis was identified by scanning the LPFM tip in the direction along the edge of the samples. 
As the sample cleaves along the crystalline axes, the scanning direction that yields the minimal LPFM amplitude is perpendicular to the polar axis and is defined as $\phi = 90\degree$.

\section{Differential Scanning Calorimeter}

Specific heat measurements were performed under a nitrogen atmosphere with a TA Instruments Q250 differential scanning calorimeter using ASTM standard E1269.
Tzero aluminum pans were weight-matched within ± 0.01 mg.
Sample, baseline, and sapphire calibrant experiments were conducted with a heating rate of 20 °C/min and 10 min isothermal segments before and after the heating ramp.
The specific heat was calculated using TA instruments TRIOS software.
At least three experiments per sample were performed, and multiple samples of 3 - 8 mg were tested.
\bibliography{main}
\clearpage

\begin{center}
\textbf{\large Data availability}
\end{center}

The raw data supporting this work is available from the corresponding authors upon request.

\begin{center}
\textbf{\large Acknowledgements}
\end{center}

We would like to thank Shireen Adenwalla, Alfred Zong, and Aaron Lindenberg for their insightful discussions.
This work is primarily supported by the National Science Foundation through EPSCoR RII Track-1: Emergent Quantum Materials and Technologies (EQUATE), award OIA-2044049.
T.L. and X.H. acknowledge the funding from the National Science Foundation via award DMR-2118828.
J.K. and L.F.B. disclose support for the research of this work from the National Science Foundation under award DMR-2437104.

\begin{center}
\textbf{\large Competing interests}
\end{center}

The authors declare no conflict of interest.
\clearpage
\begin{center}
\textbf{\large Supplementary Materials}
\end{center}

\setcounter{equation}{0}
\setcounter{figure}{0}
\setcounter{table}{0}
\setcounter{page}{1}
\setcounter{section}{0}
\makeatletter
\renewcommand{\theequation}{S\arabic{equation}}
\renewcommand{\thefigure}{S\arabic{figure}}
\renewcommand{\bibnumfmt}[1]{[S#1]}
\renewcommand{ \citenumfont}[1]{S#1}
\renewcommand{\thesection}{S-\Roman{section}}

\section{Static deflection of electron beam in TEM\label{Sec:S1}}

A quantitative estimate of the intrinsic electric field in the ferroelectric samples can be made via: 
\begin{equation}
    \gamma=\frac{e\,m^*\,t}{p^2}\,E,
\end{equation}
where $\gamma$ is the deflection angle of the electron beam, $e$ is the charge of an electron, $m^*$ and $p$ are the relativistic mass and momentum of the electrons, respectively, and $E$ is the lateral intrinsic electric field of the ferroelectric crystal. 
From the split spots in the SAED pattern, we retrieved the deflection angle to be 0.24 mrad. 
Using the relativistic mass and momentum of the 200 keV electrons in the TEM, the intrinsic electric field of NbOI\textsubscript{2} is estimated to be $7.6\times10^9$ V/m, which is on the same order of magnitude as the calculated intrinsic electric field strength using the polarization and dielectric constants from \cite{wu2022}, which is $9.0\times10^9$ V/m.
Compared to our time-resolved measurements, for a deflection angle of 0.27 mrad and the momentum of the 90 keV electrons, the calculated average change in the electric field is $1.2\times10^9$ V/m at maximum.

\section{Exclusion of domain dynamics\label{Sec:S2}}

\begin{figure}
    \centering
    \includegraphics[width=5.5 in]{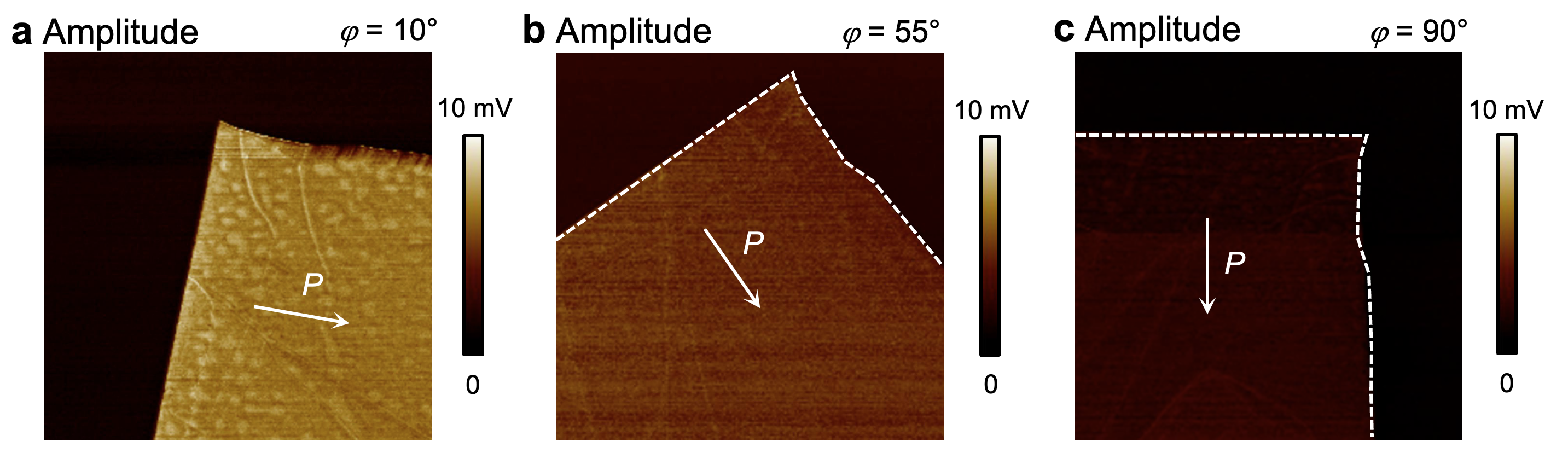}
    \caption{
    LPFM amplitude images of NbOI\textsubscript{2} flake at $\varphi$ values of \textbf{a} 10°, \textbf{b} 55°, and \textbf{c} 90°.
    The arrows mark the in-plane polarization directions.
    }
    \label{figS1}
\end{figure}

Domain dynamics may complicate the interpretation of the deflection because the diffraction pattern is spatially averaged over the entire region of interest.
The ferroelectric domain of several freshly exfoliated NbOI\textsubscript{2} samples is examined with a PFM.
Figure~\ref{fig1}c and Fig.~\ref{figS1}a show the LPFM phase and amplitude images of the flake, respectively.
We identified the polar axis of the sample by performing angle-dependent LPFM images (Methods).
As shown in Fig.~\ref{figS1}, the LPFM amplitude depends sensitively on the scanning direction with respect to the polar axis, defined as angle $\varphi$, and exhibits a minimal signal while scanning in the perpendicular direction (Fig.~\ref{figS1}c).
The LPFM images reveal a uniform polarization in the flake, showing the sample is in a single-domain state.

\par

\begin{figure}
    \centering
    \includegraphics[width=3.4 in]{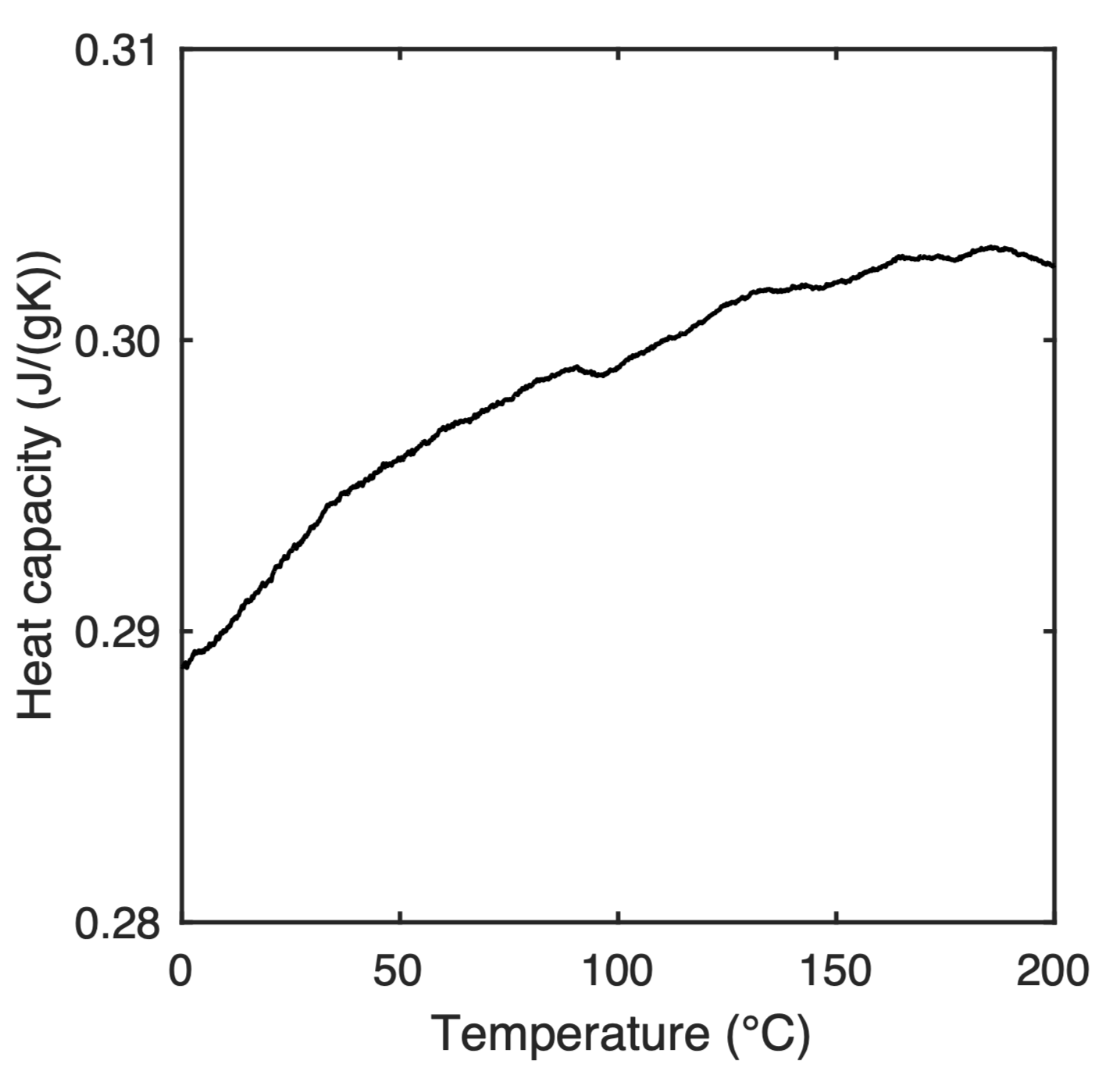}
    \caption{
    The heat capacity of NbOI\textsubscript{2} as a function of temperature. 
    }
    \label{figS2}
\end{figure}

We measured the specific heat of the NbOI\textsubscript{2} crystals using DSC. 
The temperature dependent specific heat is found to be $0.27\pm0.04$ J/(gK) at 20\degree C and $0.31\pm0.05$ J/(gK) at 80\degree C (Fig.~\ref{figS2}). 
The pump fluence is capped below 1.4 $\mathrm{mJ/cm^{2}}$, where the incident angle of the pump beam is 58\degree.
Under such optical pump fluences, the thermalized lattice temperature of the 39 nm sample is below the Curie temperature of NbOI\textsubscript{2}. 
These conditions exclude the presence of domain walls in the samples before and during the experiment.
Ultimately, pump-probe-style experiments are repetitive in nature.
Our consistent results during the experiments are strong evidence that the system relaxes to the same initial state for the hundreds of thousands of pump pulses shot at the samples. 

\section{The fluence dependence of the dynamics}

\begin{figure}
    \centering
    \includegraphics[width=5.5 in]{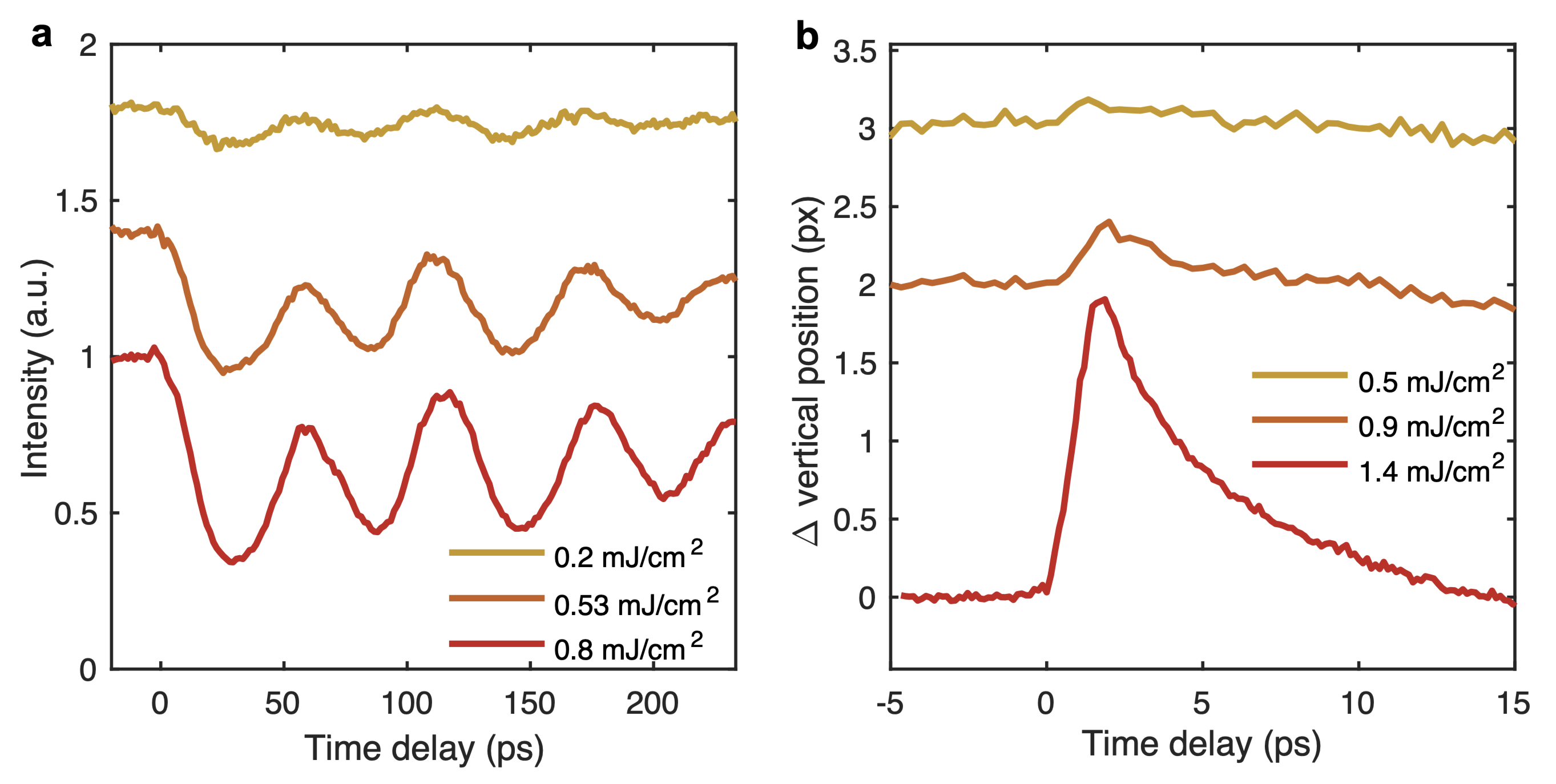}
    \caption{
    The pump fluence dependence of \textbf{a} coherent oscillational amplitude, and \textbf{b} deflection magnitude. 
    The figures are vertically displaced for clarity.
    }
    \label{figS3}
\end{figure}

When varying the fluence of the pump laser, the amplitude of the coherent acoustic waves changes accordingly (Fig.~\ref{figS3}a). 
The oscillating frequency remains the same, which is determined by the sample thickness and interlayer sound velocity. 
\par
The deflection of the electron beam under different pump fluences is displayed in Fig.~\ref{figS3}b.  
The nonlinear response of polarization reduction to the laser energy matches the nonlinear relationship between the ferroelectric order parameter and lattice temperature in a phase transition from ferroelectric state to paraelectric state. 

\section{Modes of acoustic waves\label{Sec:S4}}

There are two modes of interplane lattice vibration: the longitudinal acoustic (LA) phonon and the transverse acoustic (TA) phonon, where the displacements of the atoms are aligned with the wavevector and normal to the wavevector, respectively. 
Previously, light-induced coherent acoustics waves have been found with ultrafast electron probes \cite{park2005,harb2009,wei2017dynamic,qian2020,zong2023}. 
They usually feature a breathing motion along the thickness direction of the ultrathin samples, and less commonly, in the materials that lack inversion symmetry, the interlayer shear wave can be generated, as well \cite{zong2023}. 
We compartmentalize the two modes of the acoustic waves by tilting the sample to different zone axes. 
\par

\begin{figure}
    \centering
    \includegraphics[width=6 in]{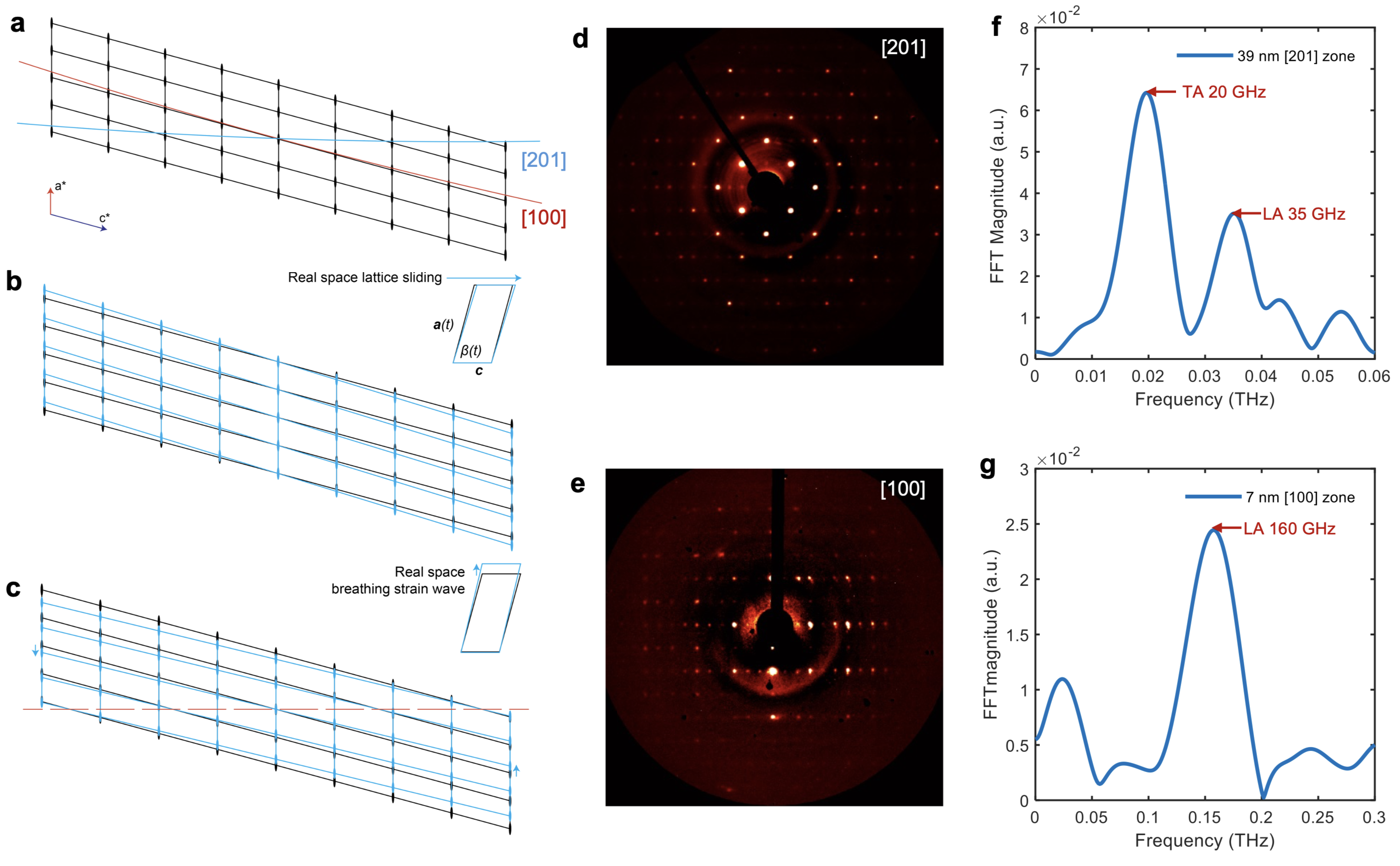}
    \caption{
    The longitudinal and transverse modes of acoustic waves. 
    \textbf{a}, The diffraction geometry is illustrated in the reciprocal space. 
    Due to the misalignment of the out-of-plane lattice vector \textbf{\textit{a}} and the surface normal, the \textbf{\textit{b\textsuperscript{*}}c\textsuperscript{*}} plane of the reciprocal lattice is tilted. 
    The blue and red curves represent the Ewald sphere of the 90 keV electrons at [201] and [100] zone axes, respectively, with an exaggerated curvature. 
    \textbf{b}, An illustration of the deformation in the reciprocal lattice due to the real-space transverse interlayer motions. 
    The inset shows the lattice sliding in real space. 
    \textbf{c}, The deformation of the reciprocal lattice due to the real-space longitudinal interlayer motions. 
    The inset shows the lattice expansion in real space.
    \textbf{d}-\textbf{e}, Electron diffraction patterns taken at [201] (d) and [100] (e) zone axes, respectively. 
    \textbf{f}-\textbf{g}, The Fourier transforms of the time-resolved diffraction intensity of ($1\overline{12}$) diffraction order taken from a 39 nm thick sample in [201] zone axis (f) and (003) diffraction order taken from a 7 nm thick sample in [100] zone axis (g), respectively.
    }
    \label{figS4}
\end{figure}

Figure~\ref{figS4}a depicts a slice of the reciprocal space geometry of the electron diffraction in the \textbf{\textit{a\textsuperscript{*}}c\textsuperscript{*}} plane that crosses the zeroth order. 
Because of the misalignment of the out-of-plane lattice vector and the surface normal, the \textbf{\textit{b\textsuperscript{*}}c\textsuperscript{*}} plane of the reciprocal lattice is tilted by 15.5\degree\ with respect to the real space \textbf{\textit{bc}} lattice plane. 
Under the influence of TA phonon, the interlayer lattice deformation features a shear motion, depicted in Fig.~\ref{figS4}b inset. 
The sliding of the lattice causes a time-variant angle $\beta(t)$ while preserving the interlayer distance. 
The resulting deformation applied to the reciprocal lattice is shown in Fig.~\ref{figS4}b, where the entire reciprocal lattice undergoes a seesaw wobbling motion. 
While the lattice is affected by the LA phonon, Fig.~\ref{figS4}c inset shows the lattice deformation of a changing interlayer distance. 
As a result of expansion along the normal direction, the reciprocal lattice will be compressed in the same direction, as shown in Fig.~\ref{figS4}c. 
In this deformed reciprocal lattice, the plane parallel to the layers in real space that crosses the zeroth order is unaffected, which is marked by the red dashed line. 
\par
In this study, we mainly take data from two Laue zones of the crystal lattice: one is the [201] zone, which is close to the surface normal; the other is the [100] zone, which is along the \textbf{\textit{a}}-axis. 
Our single-tilt sample holder allows us to rotate the sample and align the incidence of the electron pulses to the zone axes \cite{wang2023ultrafast}. 
The electron diffraction patterns without the laser excitation from [201] and [100] zone axes are partly shown in Fig.~\ref{figS4}d and e, respectively. 
The different Ewald spheres corresponding to the [201] and [100] zones are shown in Fig.~\ref{figS4}a as the blue and red curves, respectively. 
Relating to the reciprocal lattice deformations displayed in Fig.~\ref{figS4}b and c, we have: 
1) in the [201] zone axis, the Ewald sphere is almost overlapped with the red-dashed line in Fig.~\ref{figS4}c, meaning the dynamics due to the LA phonons are suppressed in this diffraction geometry, and the oscillations in the diffraction intensity are primarily a result of the TA phonons; 
2) the [100] Ewald sphere is parallel to the zeroth layer of the reciprocal lattice, and the probed dynamics are influenced by both TA and LA phonons. 
However, because the acoustic waves are generated in a thermal expansion, the LA phonons are bigger in amplitude and the primary influence in this diffraction geometry. 
\par
With [201] zone axis, the data is taken with the 39-nm-thick sample, and the frequency components in the diffraction intensity dynamics are revealed by a Fourier analysis, plotted in Fig.~\ref{figS4}f. 
In this geometry, the dominant mode of acoustic waves is the shear wave, and the prominent peak at 20 GHz clearly demonstrates its existence. 
Due to the small misalignment of the zone axis to the surface normal, the effect of the longitudinal wave is not completely suppressed, and a 35 GHz peak marks the frequency of the LA phonon corresponding to the thickness of the sample. 
The probed LA phonon in this geometry results from a slight misalignment of the [201] zone axis and the surface normal. 
From the relationship $v=2fD$, we estimate the out-of-plane sound velocity $v_L=1365\ \mathrm{m/s}\pm40\ \mathrm{m/s}$ and $v_T=780\ \mathrm{m/s}\pm40\ \mathrm{m/s}$ for the longitudinal and transverse modes, respectively, where the error comes from the uncertainty in the thickness measurements. 
In the [100] zone axis, we take data with a sample of 7 nm thickness. 
The Fourier transformation of the intensity oscillations is plotted in Fig.~\ref{figS4}g, where a single peak at 160 GHz signifies the LA phonon frequency corresponding to the 7 nm thickness. 
From it, we estimate the out-of-plane sound velocity to be $v_L=1120\ \mathrm{m/s}\pm320\ \mathrm{m/s}$, which agrees with the result we get from [201] zone axis. 
\end{document}